\documentclass[12pt]{article}

\usepackage{epsf}

\newcommand{\ket}[1]{|\; #1\; \rangle}

\newcommand{\ketp}[1]{{|\; #1\; \rangle}^{\prime}}

\newcommand{\be}{\begin{equation}}
\newcommand{\ee}{\end{equation}}
\newcommand{\ben}{\begin{eqnarray}\displaystyle}
\newcommand{\een}{\end{eqnarray}}
\newcommand{\bea}{\begin{eqnarray*}\displaystyle}
\newcommand{\eea}{\end{eqnarray*}}
\newcommand{\refb}[1]{(\ref{#1})}

\newcommand{\fn}{\footnote}
\newcommand{\sectiono}[1]{\section{#1}\setcounter{equation}{0}}

\newcommand{\bec}{\begin{center}}
\newcommand{\eec}{\end{center}}

\newcommand{\noi}{\noindent}
\newcommand{\non}{\nonumber}
\newcommand{\ul}{\underline}

\newcommand{\wt}{\widetilde}
\newcommand{\wh}{\widehat}
\newcommand{\pr}{\;\prime}
\newcommand{\ep}{\epsilon}
\newcommand{\vp}{\varphi}
\newcommand{\ssz}{\scriptsize}
\newcommand{\NSNSU}{\mbox{\ssz\NSNS,U}}
\newcommand{\NSNST}{\mbox{\ssz\NSNS,T}}
\newcommand{\RRU}{\mbox{\ssz\RR,U}}
\newcommand{\RRT}{\mbox{\ssz\RR,T}}
\newcommand{\imply}{\Rightarrow}
\newcommand{\ra}{\rightarrow}

\newcommand{\p}{\partial}
\newcommand{\x}{\times}
\newcommand{\ox}{\otimes}

\newcommand{\cpc}{\tau_2\otimes\sigma_1\otimes\Sigma_1}
\newcommand{\cpcc}{\tau_1\otimes\sigma_1\otimes\Sigma_2}

\newcommand{\y}{\sigma_1}

\newcommand{\z}{\Sigma_1}
\newcommand{\zz}{\Sigma_2}
\newcommand{\zzz}{\Sigma_3}
\newcommand{\Y}{\tau_1}
\newcommand{\YY}{\tau_2}
\newcommand{\YYY}{\tau_3}

\def\RR{R$-$R }
\def\NSNS{NS$-$NS }

\def\Zop{\bbbz}

\def\Id{\bbbi_2}                

\def\half {\frac{1}{2}}

\def\bbbz {{\sf Z\!\!Z}}

\def\bbbi {{\rm I\!I}}

\def\I{{\cal I}_4}
\def\II{{\cal I}_{4}^{\pr}}
\def\A{{\cal A}_{\vp^1}}
\def\AA{{\cal A}_{\vp^2}}
\def\T{{\cal T}}
\def\N{{\cal N}}
\def\H{{\cal H}}                 
\def\viz{{\em{viz.\ }}}
\def\ie{{\em{i.e.\ }}}
\def\bD {\bar{\rm D}}

\def\wi{(-1)^{\rm F}}                        
\def\va {\vec{a}}
\def\vb {\vec{b}}
\def\vap {\vec{a}^{\pr}}
\def\vbp {\vec{b}^{\pr}}
\def\v {\vec{0}}
\def\th {{\theta}_6}
\def\bth {\bar{\th}}
\def\thp {{{\theta}_6}^{\pr}}
\def\bthp {\bar{\thp}}
\def\tth {{\theta}_8}
\def\btth {\bar{\tth}}
\def\tthp {{{\theta}_8}^{\pr}}
\def\btthp {\bar{\tthp}}
\def\nu {{\N}_{\mbox{\ssz{U1}}}}        
                     
\def\nt{{\N}_{\mbox{\ssz{T1}}}}         
                     
\def\NU{\wt{{\N}}_{\mbox{\ssz{U0}}}}    
\def\NT{\wt{{\N}}_{\mbox{\ssz{T0}}}}    
\def\NNU{\wt{{\N}}_{\mbox{\ssz{U2}}}}   
\def\NNT{\wt{{\N}}_{\mbox{\ssz{T2}}}}   
\def\bep{\bar{\ep}}
\def\epp{{\ep}^{\pr}}
\def\bepp{\bar{\epp}}

\begin{document}

\begin{titlepage}
\thispagestyle{empty}
\begin{flushright}
{\sf hep-th/0007158\\
MRI-P-000704}\\
\end{flushright}
\vskip 1.0cm

\begin{center}
\Large{\bf Non-BPS D-branes on a Calabi-Yau Orbifold}
\vskip 0.7cm
\large{\rm Jaydeep Majumder{\fn{Email : joydeep@mri.ernet.in}} and 
Ashoke Sen{\fn{Email : asen@thwgs.cern.ch, sen@mri.ernet.in}}}
\vskip 0.3cm
{\sl Mehta Research Institute of Mathematics\\
and Mathematical Physics\\
Chhatnag Road, Jhoosi,\\
Allahabad 211019, INDIA}

\end{center}
\vskip 0.5 cm

\begin{abstract}

A system containing a pair of non-BPS D-strings of type IIA string theory
on an
orbifold,
representing a single D2-brane wrapped on a nonsupersymmetric 2-cycle of
a
Calabi-Yau 3-fold with $(h^{(1,1)},h^{(1,2)})$ = (11,11), is analyzed. In
certain region of the moduli space the configuration is stable.
We show that beyond the region of stability
the system can decay into a pair of non-BPS
D3-branes. 
At one point on the boundary of the region of stability, there
exists a marginal deformation which connects the
system of non-BPS
D-strings to the system of non-BPS D3-branes. Across any other point on
the boundary of the
region of stability, the transition from the system of non-BPS D-strings
to the system of non-BPS D3-branes is first order. We discuss the phase
diagram in the moduli space for these configurations.

\end{abstract}

\end{titlepage}

\tableofcontents


\pagestyle{plain}

\baselineskip=18pt

\sectiono{Introduction and Summary}\label{s1}

It is now well known that type IIA (IIB) string theory contains extended 
solitonic objects called D-$2p$ (D-$(2p + 1)$) branes which carry charges
of 
various Ramond-Ramond (RR) 
gauge fields present in this theory\cite{9510017}. These objects satisfy BPS 
condition and as a result can be interpreted as those solutions of classical
stringy equations of motion which preserve half of the spacetime supersymmetry
charges of the underlying theory. These objects are also amenable to a nice
dual perturbative description in that the open strings can end on them. 
Due to its BPS nature, these objects remain stable even when one goes from 
weak to strong coupling regime of string theory and thus provides a 
theoretical laboratory to test various nontrivial conjectures of string 
dualities.

As a next step towards verifying these conjectures on string dualities one 
might ask whether these dualities work for the non-BPS states present in five
different perturbative descriptions of string theory.
Generically these non-BPS
states are not stable since their mass is not protected by supersymmetry once
one goes beyond tree level of string perturbation theory. But there exists 
some stable non-BPS states simply because they are the lightest states
carrying 
some conserved quantum numbers. Such states were explored in \cite{9803194,
9805019,9805170,9806155,9808141,9809111}. As a
non-trivial 
check of type I-heterotic $SO(32)$ duality, it was shown that a stable 
state in the perturbative spectrum of the heterotic theory, which transforms 
as a spinor under the $SO(32)$ gauge group, has a dual 
counterpart in type I string theory as a non-BPS
D0-brane\cite{9808141}. Boundary state description of non-BPS D-branes 
was developed in refs.\cite{9806155,9903123}. Attempts to construct
solutions of supergravity equations of motion describing these non-BPS
D-branes have been made in refs.\cite{0003033,0003226,0005242,0007097}.

Next we can look for such stable non-BPS D-branes in compactified
theories. If we consider type IIA / IIB theory on a K3 surface or a
Calabi-Yau
3-fold, examples of BPS D-branes wrapped on supersymmetric
cycles\cite{9507158} of these compact manifolds are abundant
\cite{9603167,9704151,9712186,9808080,9906200, 9907131}. These compact
manifolds
also
contain
topologically nontrivial nonsupersymmetric cycles. These can be
interpreted as a homological sum of supersymmetric cycles. Hence one might
ask if it is possible to get stable, non-BPS configurations by wrapping a
BPS D-brane on one of these non-supersymmetric cycles. Although charge
conservation allows them to decay into two or more BPS configurations
carrying the same total charge quantum numbers, such a decay may be
prevented due to energy conservation in certain regions of the moduli
space.

Generically type IIA (IIB) theory 
contains non-BPS D-branes of odd (even) dimension. These branes are
unstable 
which is signalled by the presence of tachyon on the brane world-volume. 
However, the tachyonic mode may be projected out when we consider certain
orbifolds / orientifolds of the
theory\cite{9812031,9901014}.\footnote{Stable non-BPS D-branes on
asymmetric orbifolds have been studied recently in ref.\cite{0007126}.} In
ref.\cite{9812031} it was shown that on an orbifold representing type II
string theory compactification on K3, such a non-BPS 
D-brane represents a BPS D-brane wrapped on a nonsupersymmetric cycle and it 
can be stable in certain region
of the moduli space. Beyond the region of stability it can decay into a pair
of BPS D-branes wrapped on supersymmetric cycles. The nature and total number
of these decay products are determined by conservation laws of various quantum
numbers {\em{e.g.}} mass, bulk RR charge and twisted RR charge at various 
fixed points of the orbifold. This instability of the non-BPS
configuration is signalled by the  
reappearance of the 
tachyon on the brane world-volume. A crucial ingredient in this
analysis was that the 
tachyon at the boundary of the region of stability represents an exactly
marginal deformation, and this exactly marginal deformation interpolates
between the original system representing a D-brane wrapped on a
non-supersymmetric cycle, and the final decay product, representing a
set of D-branes, each wrapped on a supersymmetric cycle. 

In this paper we shall pursue this programme in the context of type IIA theory
on a particular Calabi-Yau 3-fold with Hodge numbers $h^{(1,1)}=11,
h^{(1,2)}=11$. In the orbifold limit this Calabi-Yau 3-fold
is obtained by modding out $T^6$ by two $\Zop_2$ transformations, whose
generators 
we shall denote by $\I$ and $\II$ respectively. We shall 
describe these transformations in detail in the next section (See
eqs.(\ref{I}),
(\ref{II}) of subsection (\ref{ss2.1})). This Calabi-Yau orbifold was discussed
in \cite{9812031,9505162,9607057,9910249} in various other contexts. In fact
certain general results of this analysis were already stated in 
ref.\cite{9812031}.
Also in ref.\cite{9910249}, it was shown that for non-BPS D-branes wrapped on
nonsupersymmetric cycles of this Calabi-Yau orbifold, the one-loop open
string
partition function vanishes at some special points of the moduli space of this
orbifold. Thus at these points in the moduli space the branes do not exert any
force on each other. This generalises the idea developed in
ref.\cite{9908060} in the context of orbifold K3.
Recently in ref.\cite{0005153} study of non-BPS branes on a different
Calabi-Yau
orbifold (the one with $(h^{(1,1)}, h^{(1,2)}) = (51,3)$) was carried out
using boundary state formalism\cite{9910109, BoundaryReview, POLCAI, CLNY,
ONOISH,
ISHIBASHI, BOUNDARYOLD, BOUNDARY} and $K$-theory
analysis\cite{9810188,9812135}.

We shall start with a $\I$ and $\II$ invariant configuration,
containing a pair of non-BPS D-strings of type IIA theory 
wrapped on one of 
the circles of $T^6$ and then mod it out by those two $\Zop_2$ transformations.
By examining the translational zero modes and the twisted sector
RR charges
carried by this configuration, one
can identify this as a single 2-brane, wrapped on a non-supersymmetric
2-cycle of the Calabi-Yau manifold. (In the orbifold limit the two cycle
collapses to a line.)  
For certain range of values of the radii of the compact directions 
the spectrum of open strings on the D-string pair does not
contain any tachyonic mode. 
Once we go beyond this region, the system develops a tachyonic mode and
becomes unstable. We show that
beyond the region of stability there is a lower energy configuration
involving a pair of non-BPS D3-branes wrapped on $T^6$,
carrying the same charge quantum numbers. Hence it is natural to
attribute the tachyonic instability of the original system to the
possibility of decaying into the new system. By examining the
translational zero modes and the charge quantum numbers, we can interprete 
the
final state as a pair of BPS D4-branes, each wrapped on a homologically
trivial 4-cycle of the Calabi-Yau manifold,\footnote{A D4-brane wrapped
on a homologically trivial 4-cycle can also be regarded as a bound state
of a D4-brane $\bD$4-brane pair, each wrapped on the same 4-cycle.} but
carrying non-zero magnetic
flux through various 2-cycles and hence carrying twisted sector RR 
charges. The
total RR charges of the initial and final configurations agree.

While comparing the energies of the initial and the final configurations,
we encounter a surprise:
in
part of the region where
the original system does not have a tachyonic mode, the new system of
D3-branes still has lower
energy.
Thus in this region of the moduli space the original system is
{\em metastable}, as it can decay into the system of D3-branes despite
being free from tachyonic modes. There is however
certain region of the moduli space where the original system of
non-BPS D-strings is the lowest energy state carrying the given charge
quantum numbers. In this region the system is absolutely {\em stable}.

The same story is repeated for the system of non-BPS D3-branes. In part of
the region
of the moduli space this system does not have a tachyon. But only in a
subspace of
this region this has mass lower than the system of D-strings, and hence is
{\em stable}. Outside this region it can decay into the system of non-BPS
D-strings and hence is {\em metastable}. As we go further out in the
moduli space, it develops a tachyonic mode and becomes {\em unstable}.

Thus we can divide
the moduli space into four regions:
\begin{enumerate}
\item D1-brane system unstable, D3-brane system stable,
\item D1-brane system stable, D3-brane system unstable,
\item D1-brane system stable, D3-brane system metastable, and
\item D1-brane system metastable, D3-brane system stable.
\end{enumerate}
There are of course other regions of the moduli space where neither of
these systems are stable, and a different configuration represents the
lowest energy state with given charge quantum numbers. But we shall focus
on the part of the moduli space spanning these four regions.

In terms of the tachyon potential, the existence of the metastable state
reflects the appearance of a local minimum of the tachyon potential,
besides the global minimum.
Recently, there has been a lot of progress in understanding the
nature of the tachyon potential\cite{9911116,9912249,0001201,0001084,
0002117,0002211,0002237,0003031,0003220,0004015,0005036} using string
field theory
\cite{SFT1, SFT2}. It will be interesting to see if string field theory
can be used to predict the existence of these local minima.

There is a special point in the moduli space where the would be
tachyonic modes on both the D1-brane system and
the D3-brane system become massless.\footnote{This is related to the point
in the
moduli space where the open string spectrum develops exact Bose-Fermi
degeneracy even though the system is non-supersymmetric\cite{9910249}.} We
show that at this point the tachyon
potential vanishes identically, and by giving vacuum expectation value
(vev) to the tachyon we can go from the D1-brane system to the
D3-brane system. In the
language of first quantized string theory, the tachyon vev represents an
exactly marginal deformation of the boundary conformal field theory (BCFT)
describing the two systems, and this deformation takes us from the BCFT
describing the pair of D-strings to the BCFT describing the pair
of D3-branes. 

The paper is organised as follows. In section \ref{s2} we discuss the 
construction of the Calabi-Yau orbifold and the configuration of D-branes on 
it. We also determine the region of (meta-)stability of the D-brane
configuration in the moduli space of
this 
orbifold. In particular, we obtain the phase diagram in the moduli space 
indicating the region of 
stability of the initial configuartion of non-BPS D-string pair and
the final 
configuration of non-BPS D3-brane pair. In section \ref{s3} we apply the
conservation 
of energy, twisted sector RR charge at various fixed points and bulk RR
charge to 
show that beyond the region of stability, the initial configuartion of
D-strings can
actually decay
into a pair of non-BPS D3-branes.\footnote{To avoid confusion, we mention
that this (and only this) part of the
analysis 
is done using
an
equivalent type IIB language.} We also argue that each of these non-BPS
D3-branes can be regarded as a BPS D4-brane wrapped on a homologically
trivial 4-cycle of the Calabi-Yau manifold, carrying non-zero magnetic
flux through different 2-cycles. In section \ref{s4} we show
that at the ``critical radii'', which represent a 
special point on the boundary of the region of stability, some particular 
tachyonic mode
becomes exactly marginal. Following ref.\cite{0003124,0005114} we study the
effect of switching on this marginal deformation on the open string
spectrum, and show that it interpolates between the BCFT's describing the 
D1-brane pair and the D3-brane pair.
This establishes that the possible decay mode identified in section
\ref{s3} is the actual decay mode of the system beyond the region of
stability. 

\sectiono{Non-BPS D-strings on Calabi-Yau Orbifold}\label{s2}

In this section we shall describe the system we are going to study.
We shall consider non-BPS D-string of type IIA theory in a particular 
Calabi-Yau orbifold background. The relevant Hodge number for this Calabi-Yau
3-fold is $(h^{(1,1)},h^{(1,2)}) = (11,11)$. We begin by describing the
constuction of this 
particular orbifold.

\subsection{ Construction of the Calabi-Yau Orbifold}\label{ss2.1}

We begin with a six 
dimensional torus, $T^6$ and label its coordinates by $X^4,X^5,\cdots ,X^9$. 
We shall focus on the subspace of the moduli space where it can
be represented as a product of six circles. Let $R_4, R_5, \cdots , R_9$ denote
the radii of these six compact directions. Now we mod out this $T^6$ by two 
$\Zop_2$ symmetries, generated by $\I,\; \II$,

\be\label{I}
 \I :  \quad\left\{
\begin{array}{rcr}
(X^4, X^5,X^6,X^7,X^8,X^9) & \longrightarrow &  (X^4,X^5,-\;X^6,-\;X^7,-\;
X^8,-\;X^9)\non\\
(\psi^4,\psi^5,\psi^6,\psi^7,\psi^8,\psi^9) & \longrightarrow &  
(\psi^4,\psi^5,-\;\psi^6,-\;\psi^7,-\;\psi^8,-\;\psi^9)
\end{array}\right.\;.  
\ee
and
\be\label{II}
\II :\quad\left\{
\begin{array}{rcr}
X^4 & \longrightarrow & -X^4\\ X^5 & \longrightarrow & -X^5\\
X^6 & \longrightarrow & -\;X^6 + \pi R_6\\ 
X^7 & \longrightarrow & -\;X^7\\
X^8 & \longrightarrow & X^8 + \pi R_8\\ 
X^9 & \longrightarrow & X^9\\[1mm]
\psi^{4,5,6,7} & \longrightarrow & -\;\psi^{4,5,6,7}\\
\psi^{8,9} & \longrightarrow & \psi^{8,9}
\end{array}\right.\;.
\ee 
where $\psi^4,\psi^5, \cdots , \psi^9$ are the worldsheet fermions
corresponding to
those six directions. The action of $\I,\II$ on these worldsheet fermions
has been determined from the condition that these two discrete symmetries
do not break worldsheet supersymmetry (i.e. both $\I$ and $\II$ should
commute 
with the worldsheet supercurrent, $T_F$). Both $\I$ and $\II$ leave the
non-compact coordinates $X^0,\ldots X^3$ and their fermionic partners
invariant.

\renewcommand{\arraystretch}{1.3}
\begin{table}[!ht]\label{actionII}
\bec
\begin{tabular}{||c|c|c|c|c|c|c||}\hline
  & $X^4$ & $X^5$ & $X^6$ & $X^7$ & $X^8$ & $X^9$ \\[1.5mm]\hline
$\I$ & + & + & $-$ & $-$ & $-$ & $-$ \\[1.5mm]\hline
$\II$ & $-$ & $-$ & $-$,$\;\half$ & $-$ & +,$\;\half$ & + \\[1.5mm]\hline
$\I\II$ & $-$ & $-$ & +,$\;\half$ & + & $-$,$\;\half$ & $-$ \\[1.5mm]
\hline
\end{tabular}
\eec
\caption{Summary of the action of $\I$ and $\II$ on $T^6$.}
\end{table}

In an obvious notation, the equations (\ref{I}) and (\ref{II}) can be 
summarised as shown in table 1.
We shall denote this Calabi-Yau orbifold as $T^6/\Zop^{(1)}_2\otimes
\Zop^{(2)}_2$;
in particular it can be shown that it has $(h^{(1,1)},h^{(1,2)}) = (11,11)$ 
\cite{9505162}. This particular background preserves only 1/4th. of the
space-time
supersymmetry. 


\subsection{The Configuration}\label{ss2.2}

\begin{figure}[!t]  
\begin{center}
\leavevmode
\hbox{%
\epsfxsize=5.0in
\epsfysize=2.5in
\epsffile{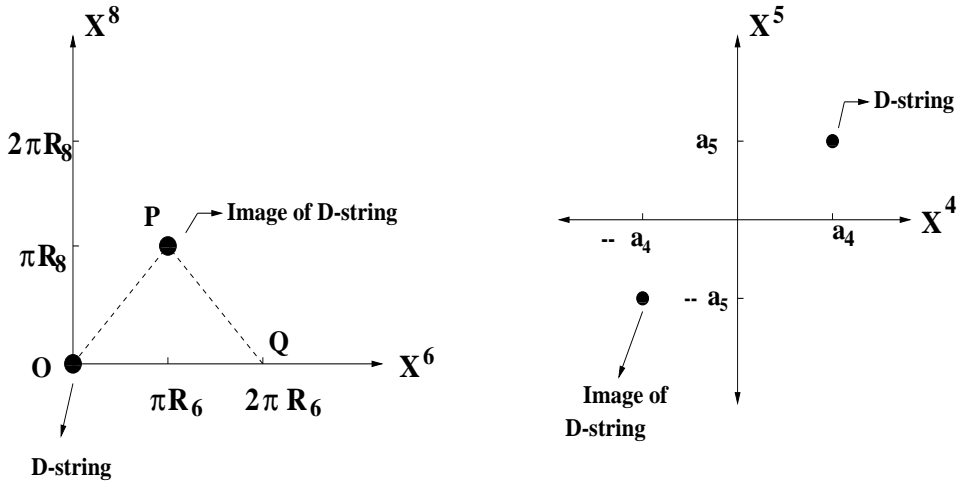}}
\caption{Location of Non-BPS D-string and its $\II$-image in the $X^{6}$-
$X^{8}$ and $X^4$-$X^5$ plane. D-strings are also present at the images
of $O$ under translation by $2\pi R_i$ along $X^i$, {\em e.g.} at the
point $Q$.} \label{f1}
\end{center}
\end{figure}

We shall consider type IIA theory on this particular Calabi-Yau orbifold
$T^6/\Zop^{(1)}_2\otimes \Zop^{(2)}_2$. We take a non-BPS D1-brane of 
this theory and wrap it along $X^9$ such that it stretches over a 
fundamental interval $[0,2\pi R_9]$. The other eight spatial coordinates
of 
this D-string are as given below :
$$
(X^1, X^2, X^3, X^4, X^5, X^6, X^7, X^8) = (b_1, b_2, b_3, a_4, a_5,
0,0,0)\, ,
$$
where $a_i$ and $b_k$ are arbitrary numbers.
Clearly the above configuration is invariant
under $\I$
but not under $\II$. The image of the D-string under $\II$ is located at
$$
(X^1, X^2, X^3, X^4, X^5, X^6, X^7, X^8) = (b_1, b_2, b_3, -a_4, -a_5, \pi
R_6,0,\pi R_8)\, .
$$
Thus the original D-string together with its $\II$-image constitute an 
$\I\ox\II$-invariant configuration (See Fig.\ref{f1}). Since the
coordinates of the D-string and its $\II$ image in the non-compact
directions $X^1,X^2,X^3$ must be identical, we see that this represents a
single object in the orbifold theory. The physical interpretation of this
object can be found by examining its charge quantum numbers. As is well
known, a non-BPS D-string of type IIA string theory does not carry any
bulk RR charge. However, if a D-string passes through an orbifold fixed
point then it carries twisted sector RR charge associated with that fixed
point\cite{9806155,9812031}. In the language of Calabi-Yau manifold,
twisted sector RR charges are carried by BPS D2-branes wrapped on the 
(collapsed)
2-cycle associated with that given fixed point. Thus the D-string system,
after
orbifolding, carries charges corresponding to BPS 2-branes wrapped on
various 2-cycles. Since it describes a single object, and is non-BPS, the
natural interpretation of this object is that it describes a single BPS
2-brane, wrapped on a non-supersymmetric 2-cycle of the Calabi-Yau
manifold. In the orbifold limit, this 2-cycle (of minimal area) collapses
to a line.

\subsection{Region of Stability in Moduli Space}\label{ss2.3}

Next we are going to determine the region of stability in the moduli space
for this particular system. Throughout our discussion we shall restrict
ourselves to tree level of open string theory. 
Also, to simplify the discussion, in this subsection we shall work on the
4-fold cover of
the orbifold, namely on the original torus $T^6$, although we shall ensure
that
we always work with configurations which are $\I$ and $\II$ invariant.
Thus the various mass formul\ae\, we shall be writing down will refer to
the
masses of the corresponding systems on $T^6$ {\it before orbifold
projection}.

Typically, the instability
of the system arises due to open strings stretched between the D-string
and one of its images developing a tachyonic mode. This involves images of
the D-string under $\II$, as well as under translation by $2\pi R_i$
along $X^i$ for ($4\le i\le 8$). Since the physics of the instability
arising due to open string stretched between the D-string and its image
under translation is identical to that in the case of K3 orbifold, and can
be analysed in a manner similar to that discussed in ref.\cite{9812031},
we shall concentrate on the instability
arising due to the open string stretched between the D-string and its
$\II$ image. 
In this case the relevant part of the moduli space is 
effectively two dimensional, spanned by $R_6$ and $R_8$, controlling the
distance between the 
D-string and its $\II$-image. The open strings stretched between the
D-string and its $\II$ image 
have been denoted as ${OP}$ and ${PQ}$ in Fig.\ref{f1}. 

The general strategy for this kind of analysis has been stated in 
\cite{0003124}. As a first step, let us determine the effective 
$(mass)^2$ of the winding mode tachyon coming from the open string ${OP}$ 
and ${PQ}$. From Fig. \ref{f1} we see that it is given by\fn{ The 
subscript $T$ and $(1)$ in the left hand side of the formula
(\ref{tachmass}) stand for
`tachyon' and D1-brane respectively.} :

\be\label{tachmass}
m_{T(1)}^2 = \Big(\frac{a_4}{\pi}\Big)^2 + \Big(\frac{a_5}{\pi}\Big)^2 + 
\Big( \half R_6\Big)^2 +\Big(\half R_8\Big)^2 
- \half
\ee
in $\alpha'=1$ unit which we shall be using throughout this paper.
Since $a_4$ and $a_5$ are
degress of freedom of the D-strings one requires that there should be no 
tachyons for any values of $a_4$ and $a_5$. The most stringent condition
comes from
\be\label{a4=a5}
a_4 = \; a_5 =\; 0\, .
\ee
Then the condition for the absence of tachyon from
the winding modes ($m_{T(1)}^2 \ge 0$) is given by :
\be\label{critical1}
R_6^2 + R_8^2 \ge 2\, .
\ee
The {\em{critical curve}} in the
$R_6$-$R_8$ plane, where the tachyon is massless, is given by 
\be\label{critical2}
R_6^2 + R_8^2 \; = \; 2\, .
\ee 

In the next section (section \ref{s3}) we shall find the possible decay 
products of these $\II$-invariant pair of non-BPS D-strings by applying the 
conservation of energy and bulk and twisted RR charges. 
Anticipating the results of the section \ref{s3}, $-$ that the instability
of 
pair of non-BPS D-strings for $R_6^2 + R_8^2  <  2$ represents the 
possibility of decay into a pair of $\II$-invariant non-BPS D3-branes 
extended in $X^6,X^8$ and $X^9$ directions, $-$ we shall now determine the 
phase diagram of this system on $R_6$-$R_8$ plane. 
As we shall see in section \ref{s3}, conservation of twisted sector RR
charge (after orbifolding) requires that on one of the
non-BPS D3-branes there are $\Zop_2$ Wilson lines along $X^6$ and $X^8$
directions.\fn{Before modding out the system by $\Zop_2^{(1)}\ox
\Zop_2^{(2)}$ transformation, the pair of non-BPS D-strings and the pair
of non-BPS 
D3-branes are actually related to each other by two $T$-dualities along $X^6$
and $X^8$.} The zero momentum mode of the open string with both ends on
the same D3-brane is projected out by requiring invariance under $\I$, and
the potentially tachyonic mode on this system comes from open
strings stretched between the pair of D3-branes 
carrying half
unit of momenta along $X^6$ and $X^8$ directions.  The mass of
this mode when the two D3-branes coincide in the $X^1,\ldots X^5,X^7$ 
directions, is given by: 
\be\label{tachmass3} 
m_{T(3)}^2 = \Big(\frac{1}{2R_6}\Big)^2 + 
\Big(\frac{1}{2R_8}\Big)^2 - \half 
\ee
Thus the condition for {\em{absence of}} tachyonic modes on the
D3-brane system is:
\be\label{tachmass4}
\frac{1}{R_6^2} +\frac{1}{R_8^2}\; \ge\;\; 2 
\ee
\begin{figure}[!t]  
\begin{center}
\leavevmode
\hbox{
\epsfxsize= 3.8in
\epsfysize= 5.8in
\epsffile{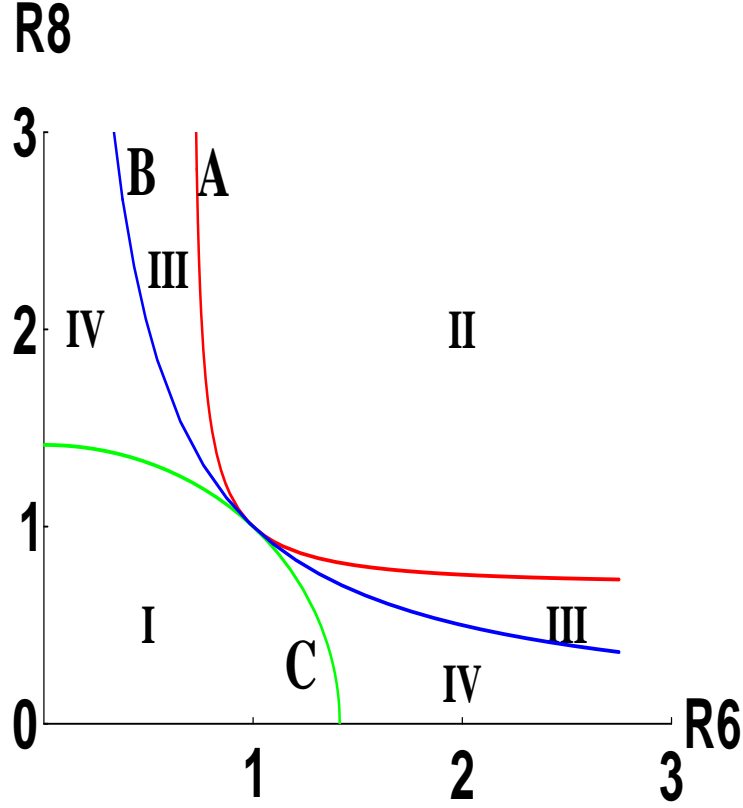}}
\caption{Phase diagram on $R^6$-$R^8$ plane. $A$ denotes the curve $R_6^{-2} +
R_8^{-2} = 2$; $B$ denotes the curve $R_6R_8 = 1$ and $C$ represents the curve
$R^2_6 + R^2_8 = 2$. In region {\bf I} the D3-brane pair is stable and the
D1-brane pair is unstable whereas in 
region {\bf II} D1-brane pair is stable and the D3-brane pair is unstable.
In region {\bf III} the D1-brane pair is stable and the D3-brane pair is 
metastable, whereas in region {\bf IV} the D3-brane pair is stable 
and the D1-brane pair is metastable.}
\label{f2}
\end{center}
\end{figure}

Let us further write down the mass formula for the
pair of D1-branes as well as for the pair of D3-branes. The mass of a
pair of non-BPS D-strings
wrapped along $X^9$ is given by:
\be\label{massD1}
M_{D1} = 2\cdot\frac{1}{2\pi g}\sqrt{2}\cdot2\pi R_9
= \frac{2\sqrt{2}}{g}R_9
\ee
where $g$ is the closed string coupling constant and the factor of $\sqrt{2}$ 
is due to the non-BPS nature of the D-string. Similarly the mass of the
non-BPS 
D3-brane pair is given by:
\be\label{massD3}
M_{D3} = 2\cdot\frac{1}{(2\pi)^{3}g}\cdot\sqrt{2}\cdot(2\pi)^{3}
R_{6}R_{8}R_{9}
= \frac{2\sqrt{2}R_{6}R_{8}R_{9}}{g}
\ee
Comparing eqs.\refb{massD1} and \refb{massD3} we see that
\be \label{ecomp}
M_{D1}\le M_{D3} \qquad \hbox{for} \qquad R_6 R_8 \ge 1\, .
\ee
Note that 
for $R_6R_8 = 1$, these two masses become degenerate. 

The various curves found during the analysis have been plotted in the
phase diagram of Fig. \ref{f2}.
The curve $R_6R_8 = 1$ is the curve $B$ in Fig. \ref{f2}. 
The equality in (\ref{tachmass4}) has been represented as the curve $A$,
and the curve $C$ denotes the circle given
by eq.(\ref{critical2}). 
There are four regions of interest altogether in this phase diagram:
\begin{enumerate}
\item $R_6^2 + R_8^2 < 2$  (Region {\bf{I}})


Here 
$$ R_6^{-2} + R_8^{-2} = {4 (R_6^2 + R_8^2) \over (R_6^2+R_8^2)^2 -
(R_6^2 - R_8^2)^2} > 2\, . $$
Thus the D3-brane pair does not have tachyon. Furthermore 
$$
R_6R_8 = \displaystyle{{R^2_6 + R^2_8 - (R_6 - R_8)^2\over 2} < 1}\;,
$$ 
and
hence the D3-brane pair is lighter than the D1-brane pair. Thus the D3-brane
pair is {\em{stable}}. On the other hand in this region the D-string pair
has tachyon and hence is {\em{unstable}}.

\item $R_6^{-2} + R_8^{-2} < 2$  (Region {\bf{II}})


Here 
$$ R_6^{2} + R_8^{2} = {4 (R_6^{-2} + R_8^{-2}) \over
(R_6^{-2}+R_8^{-2})^2 -
(R_6^{-2} - R_8^{-2})^2} > 2\, . $$
Thus the D-string pair does not have tachyon. Furthermore
$$
R_6R_8 = \displaystyle{{2\over R^{-2}_6 + R^{-2}_8 - (R_6^{-1} - 
R_8^{-1})^2} > 1}\;,
$$
and hence the D-string pair is lighter than the D3-brane pair. Thus the 
D-string pair is {\em{stable}}. But in this region
the D3-brane pair
has a tachyonic mode and hence is {\em{unstable}}.

\item $R_6R_8 > 1$ and $R_6^{-2} + R_8^{-2} > 2$ (Region
{\bf{III}})


Here $R_6^2+R_8^2 = (R_6^{-2}+R_8^{-2}) R_6^2 R_8^2 > 2$. Thus
both D1 and D3-brane pairs are free from tachyonic mode. But
$M_{D3} > M_{D1}$. 
Thus the D-string pair is stable, whereas the D3-brane pair is in a 
{\em{metastable}} state.

\item $R_6R_8 < 1$ and $R_6^{2} + R_8^{2} > 2$ (Region {\bf{IV}})


Here $R_6^{-2}+R_8^{-2}=(R_6^2+R_8^2)/(R_6^2R_8^2) > 2$. Thus
in this region also both systems are free of tachyon. But now 
$M_{D1} > M_{D3}$. Therefore the D3-brane pair is stable and the D-string
pair 
is in a {\em{metastable}} state.
\end{enumerate}


Interestingly all three curves $A$, $B$, $C$ meet at one point {\viz} 
at the {\em{critical radii}} $(R_6,R_8) = (1,1)$. 
As we shall see in section \ref{s4}, at this point the
D-string pair can be deformed to the D3-brane pair via a marginal
deformation. Across any other point along the curve $B$, the transition 
from one to the other is first order (metastable $\to$ stable). As 
neither system contains a relevant perturbation, it is not clear if the 
arguments based on renormalization group flow\cite{9406125,0003101} can 
be used to describe this transition.

Before we end this section, we should remind the reader that in our
analysis we have ignored the possibility of tachyons appearing in various
other ways. Thus for example, requiring that there is no tachyonic mode on
the D1-brane system from open strings stretched between a D1-brane and its
image under translation along $X^i$ ($6\le i\le 8$) gives the 
constraints\cite{9812031}:
\be \label{eother}
R_i \ge {1 \over \sqrt 2} \, , \qquad 6\le i \le 8\, .
\ee
On the other hand, requiring that open string states with both ends on the
same D-string, but carrying momentum along $X^9$ does not give rise to a
tachyonic mode gives
\be \label{eother2}
R_9 \le \sqrt{2}\, .
\ee
As shown in ref.\cite{9812031}, beyond these ranges of $R_i$ the system
becomes unstable against decay into a pair of BPS states.
Throughout our analysis we shall work in regions of the moduli space where
eqs.\refb{eother} and \refb{eother2} are satisfied.

Now we turn our attention to various conservation laws to verify that the
decay modes described above are indeed possible.


\sectiono{ Determination of Decay Products of Non-BPS D-string Pair
}
\label{s3}

In this section we shall show that a pair
of non-BPS D-strings as shown in Fig.~\ref{f1} (with $a_4 = a_5 = 0$) can 
decay into a pair of non-BPS D3-branes for $R_6R_8<1$, satisfying
the conservation of energy, bulk RR-charge and twisted
RR-charge. Then we shall find a physical interpretation of the final state
D3-branes as wrapped D-branes on the Calabi-Yau 3-fold.

\subsection{Verification of the Conservation Laws} \label{s3.1}

In this section we 
shall map our problem to an equivalent type IIB description to make 
calculations simpler. For this we shall perform a T-duality transformation
${\T}_9$ along the coordinate $X^9$ to get type IIB string theory on a
dual torus
labelled by $X^4,\ldots X^8, \wt X^9$.
In that case,
$$
\mbox{IIA on $T^6/\I\ox\II$}\;\; \stackrel{\T_9}{\longrightarrow}\;\; 
\mbox{IIB on $T^6/(\wt\I \cdot (-1)^{F_L} \ox\II)$}\non
$$
where $\wt\I$ denotes the transformation:
$$ \wt\I: \qquad (X^4,\ldots X^8, \wt X^9)\to (X^4, X^5, -X^6, -X^7, -X^8,
-\wt X^9)\, ,
$$
and $(-1)^{F_L}$ changes the sign of all the (R,NS) and (R,R) sector states.
We shall continue to denote by $\I$ the transformation $\wt\I\cdot
(-1)^{F_L}$ in the type IIB string theory, as this is just the image of
the transformation $\I$ in the type IIA theory. Also under the duality
${\T}_9$,
$$ \mbox{Non-BPS
D1-brane (along $X^9$) of IIA}\;\;
\stackrel{\T_9} {\longrightarrow}\;\; \mbox{Non-BPS D0-brane of IIB}
$$
$$ \mbox{Non-BPS
D3-brane (along $X^6$-$X^8$-$X^9$) of IIA}\;\;
\stackrel{\T_9} {\longrightarrow}\;\; \mbox{Non-BPS D2-brane (along 
$X^6$-$X^8$) of IIB}
$$

\begin{figure}[!th]  
\begin{center}
\leavevmode
\hbox{%
\epsfxsize=2.0in
\epsfysize=1.5in
\epsffile{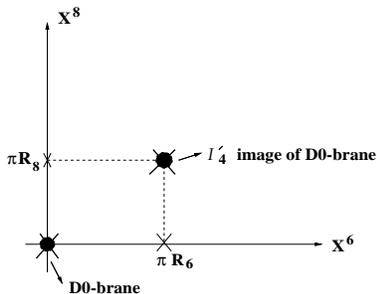}}
\caption{Non-BPS D0-brane of type IIB and its $\II$-image in the interval 
$0\le x^6< 2\pi R_6$, $0\le x^8 < 2\pi R_8$. $`\x$' denotes the
fixed points of $\I$ on $X^6$-$X^8$ plane.}\label{f3}
\end{center}
\end{figure}
Thus, in IIB description, the initial system consists of non-BPS D0-branes
at $(X^6,X^8) = (0,0)$ 
and $(\pi R_6,\pi R_8)$ (see Fig.\ref{f3}).
{}From now on, unless otherwise stated, we shall use the IIB description
throughout this section \ref{s3.1}. 
In applying conservation laws of RR charges, we shall find it convenient
to work with a double cover of the system, where we have modded out type
IIB string theory on $T^6$
by $\I$ but not by $\II$. Since $\II$ does not have any
fixed
point, and hence does not introduce any new massless state and twisted
sector RR charge,
ensuring conservation of RR charge before $\II$ modding is sufficient to
guarantee its conservation after $\II$ modding. On the other hand for
applying conservation of energy, we shall work with the four fold cover of
the system, $-$  on the original torus $T^6$. Of course, at every stage
of the analysis we
shall ensure that we have a system that is invariant under both $\I$ and
$\II$, so that we can mod out the configuration by these transformations
at the end.

\begin{figure}[!thb]  
\begin{center}
\leavevmode
\hbox{%
\epsfxsize=3.0in
\epsfysize=1.5in
\epsffile{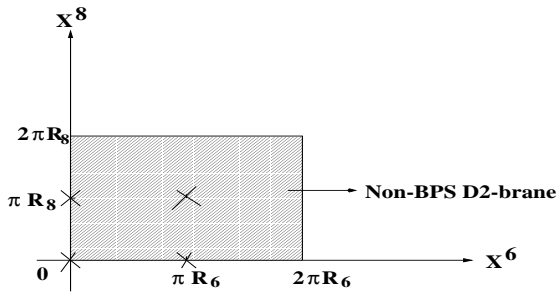}}
\caption{Non-BPS D2-brane of type IIB in the interval 
$0\le x^6< 2\pi R_6$, $0\le x^8 < 2\pi R_8$. `$\x$' denotes
the
fixed points of $\I$.}\label{f5}
\end{center}
\end{figure}
The non-BPS D0-branes, situated at (0,0) and $(\pi R_6, \pi R_8)$, do not
carry any bulk RR charge, but carry
twisted sector RR charge\cite{9806155}. 
We shall denote
by $q_{(a_6,a_8)}$ the twisted sector charge at the fixed point
$(a_6,a_8)$ in the $(X^6, X^8)$ plane. The charge will be normalized such
that a non-BPS D0-brane of IIB, placed at any of the fixed points of
$\I$, carries unit twisted sector RR charge associated with that fixed
point.
Thus our initial state has charges:
\be\label{econdtwo}
q_{(0,0)}^{(in)}=q_{(\pi R_6, \pi
R_8)}^{(in)} = 1, \qquad 
q_{(0,\pi R_8)}^{(in)}=q_{(\pi R_6, 0)}^{(in)} = 0 \, .
\ee
The final decay product must
also carry this charge. We shall consider non-BPS D2-branes of type IIB
theory in the $X^6-X^8$ plane as our
candidate for decay product. 
Fig.\ref{f5} shows the configuration of a single non-BPS D2-brane in the
range $0\le x^6< 2\pi R_6$, $0\le x^8< 2\pi R_8$. This is
by itself
an $\II$-invariant configuration. Although this does not carry any bulk RR
charge, it carries twisted sector RR charges associated with every fixed
point that it passes through. These charges can be computed using the
boundary state formalism of ref.\cite{9910109}.
If $\theta_6$ and
$\theta_8$ denote the Wilson lines on the 2-brane along $X^6$ and $X^8$
respectively, taking values 0 or $\pi$ due to the requirement of $\I$ 
invariance, then the twisted sector charges carried by the brane is given by:
\be \label{etbrch}
q_{(0,0)}={1\over 2} \ep, \qquad q_{(0,\pi R_8)} = {1\over 2} \ep
e^{i\theta_8}, \qquad q_{(\pi R_6, 0)} = {1\over 2} \ep e^{i\theta_6},
\qquad q_{(\pi R_6, \pi R_8)} = {1\over 2} \ep 
e^{i(\theta_6+\theta_8)}\, ,
\ee 
where $\ep$ can take values $\pm 1$. This result, including the overall
normalization factor of $(1/2)$ appearing in the expression for the
charges, can be read out from the expression for the boundary state of the
2-brane as given in ref.\cite{9910109}. This normalization factor of (1/2)
can also be derived from the fact that in type IIB string theory on
$T^6/\I$, the D2-brane can be continuously deformed, via tachyon
condensation, to a D-string at $X^6=0$, and a $\bD$-string at $X^6=\pi
R_6$, both stretched along $x^8$\cite{9812031}. Each of these D-strings
carry half unit of
twisted sector charge
at the fixed points through which they pass\cite{9805019,9906109}. Since a
continuous deformation cannot change the value of RR charge quantum
numbers, the D2-brane before tachyon condensation must also carry half
unit of twisted sector charge at each of the four fixed points (0,0),
$(0,\pi R_8)$, $(\pi R_6, 0)$ and $(\pi R_6, \pi R_8)$. 

Since $\II$
exchanges the points (0,0) with $(\pi R_6, \pi R_8)$ and $(0, \pi R_8)$
with $(\pi R_6, 0)$, we see from \refb{etbrch} that in order to get an
$\II$ invariant configuration, we must have $\theta_6=\theta_8=0$ or
$\theta_6=\theta_8=\pi$.

Now let us take $N$ such coincident D2-branes. We shall slightly
generalise
our notation by replacing $(\ep,\;\th,\;
\tth)$ in the above expressions by $(\ep^{(k)},\;\th^{(k)},\;\tth^{(k)})$
respectively ($k = 1, \cdots , N$). Thus for $N$ such coincident branes we 
have:
\ben\label{tw9}
q^{(tot)}_{(0,0)} & = & {1\over 2} \sum_{k = 1}^{N} \ep^{(k)}\non\\
q^{(tot)}_{(0,\pi R_8)} & = &{1\over 2} \sum_{k = 1}^{N}\ep^{(k)}
e^{i\,\tth^{(k)}}\non\\
q^{(tot)}_{(\pi R_6,\pi R_8)} & = & {1\over 2} \sum_{k = 1}^{N}\ep^{(k)}
e^{i\,(\th^{(k)}\;+\;\tth^{(k)})}\non\\
q^{(tot)}_{(\pi R_6,0)} & = & {1\over 2} \sum_{k = 1}^{N}\ep^{(k)}
e^{i\,\th^{(k)}}
\een
Again, $\II$ invariance of individual branes requires
$\theta_6^{(k)}=\theta^{(k)}_8=0$ or $\pi$ for each $k$.
Hence applying charge conservation, and using eqs.\refb{econdtwo} and 
\refb{tw9}, we get: 
\ben
\sum_{k = 1}^{N}\ep^{(k)}\,e^{i\,\tth^{(k)}} & = & 0\label{tw12}\\
\sum_{k = 1}^{N}\ep^{(k)}\,e^{i\,\th^{(k)}} & = & 0\label{tw13}\\
\sum_{k = 1}^{N} \ep^{(k)} & = & 2 \label{tw14}\\
\sum_{k = 1}^{N}\ep^{(k)}\,e^{i\,(\th^{(k)}\;+\;\tth^{(k)})} & = &  
2 \label{tw15}
\een 
Since $\ep^{(k)} = \pm 1$, from (\ref{tw14}) we can easily see that $N$
must
be an even integer.
It is also easy to demonstrate that eqs.\refb{tw12}-\refb{tw15} admit
solutions for any even integer $N$. For example, for $N=2$ we get the
following consistent set of solutions:
\be \label{econ0}
\ep^{(1)} = \ep^{(2)} = +\,1\, ,
\ee
\be\label{consistent1}
\mbox{Either}\qquad \th^{(1)} = 0,\;\; \tth^{(1)} = 0;\qquad \th^{(2)} = 
\pi,\;\; \tth^{(2)} = \pi
\ee
\be\label{consistent2}
\mbox{or}\qquad \th^{(1)} = \pi,\;\; \tth^{(1)} = \pi; \qquad \th^{(2)} = 0,
\;\; \tth^{(2)} = 0
\ee
As is evident from (\ref{consistent1}) and (\ref{consistent2}), one of
the 2-branes
carries $\Zop_2$ Wilson lines along $X^6$ and $X^8$.
{}From this solution, we can construct solutions for any even $N$ by
adding pairs of
branes with same $\theta_6$ and $\theta_8$ and opposite values of $\ep$.
Other solutions are also possible.

Now we want to compare the masses of the initial and the final system of
D-branes.
Before any orbifold projection,
the mass of $N = 2\,m\, (m\in \Zop_{+})$ non-BPS D2-branes in a
fundamental
cell
$X^6 \in [0,2\pi\,R_6]$, $X^8 \in [0,2\pi\,R_8]$ (see fig.~\ref{f5}) is
equal to
$\displaystyle\frac{N\,\sqrt{2}\,R_6\,R_8}{g}$, whereas the mass of the 
initial configuration of a pair of non-BPS D0-branes is 
$\displaystyle\frac{2\,\sqrt{2}}{g}$. Demanding that the initial system 
has higher mass than the final state, we 
obtain 
\be\label{massequal}
R_6\,R_8 \le \frac{2}{N},
\ee
This condition is least stringent for $N=2$, and hence 
this is the generic
decay channel of the
original system in its region of instability. 
Indeed, for $N=2$, eq.\refb{massequal} is automatically satisfied whenever
$R_6^2+R_8^2\le 2$, $-$ the region where the original D0-brane pair
develops a tachyonic mode. In addition, in more
restricted regions of the moduli space, given by eq.\refb{massequal} for
$N\ge 4$, additional decay channels open up in which the original system
decays to four or more non-BPS D-branes. We  should of course
keep in mind that once one
of the radii is sufficiently small, we violate \refb{eother}, and
decay
channels into BPS branes also open up.

For $N = 2$, the equality in (\ref{massequal}) and eq.(\ref{critical2})
are satisfied
simultaneously for 
$(R_6,R_8) = (1,1)$. As we shall see in the next section, there is an
exactly marginal deformation which interpolates between the initial and
the
final system at this critical radii for $N=2$. 

\subsection{Interpretation of the Final State}

So far we have analysed the system before modding it out by $\II$. As has
already been noted, after modding out by $\II$, the initial system
describes a single brane configuration as it has only one set of degrees
of freedom describing its movement in the non-compact directions. In
contrast, each of the type IIB D2-brane configurations in the final state
is $\I$
and $\II$ invariant by itself, and hence can move in the non-compact
directions independently of the other D-branes. Thus the final state
describes a set of $N$ independent objects. 

In order to find the physical interpretation of these objects, we examine
the RR charges carried by each of these objects. For this we go back to
the type IIA description by performing T-duality along $X^9$. Under this
duality, a non-BPS D2-brane of IIB gets mapped to a non-BPS D3-brane of
IIA
spanning the $X^6-X^8-X^9$ directions. The brane does not carry any bulk
RR charge, but it carries twisted sector RR charges associated with each
fixed point through which it passes. Since these RR charges are carried by
BPS
2-branes wrapped on homologically non-trivial 2-cycles, it will be natural
to interprete the non-BPS 3-brane as a 2-brane wrapped on some complicated
2-cycle of the Calabi-Yau manifold. However this interpretation cannot be
correct. To see this, let us deform all the radii $R_4,\ldots R_9$
to very large values. This makes the non-BPS 3-brane unstable, but it
continues to exist as a classical solution of the equations of motion.
Clearly in this region this cannot be interpreted as a BPS 2-brane wrapped
on
a 2-cycle, since it occupies a 3-dimensional subspace. Instead it can be 
interpreted as
a bound
state
of a D4-brane $\bD$4-brane pair\cite{9805019,9808141}, each wrapped on a 
K3 subspace of the Calabi-Yau orbifold spanned by $X^6,\ldots X^9$. The 
two D4-branes must
carrying different amounts of magnetic flux through the homology two 
cycles, so that after combining the effect of
the magnetic flux with the effect of the non-trivial anti-symmetric tensor
field flux through the 2-cycles\cite{9507012}, we recover the correct RR
charges
of the brane.
Equivalently, we can think of each non-BPS 3-brane as a single D4-brane 
wrapped on a trivial 4-cycle, but carrying nontrivial
magnetic flux
through various  homology 2-cycles.

\sectiono{Conformal Field Theory at the Critical Radii}\label{s4}

In this section we shall show that at the critical radii $R_6=R_8=1$,
there is an exactly marginal deformation which interpolates between the
D3-brane pair and the D1-brane pair. We shall carry out this analysis
before either $\I$ or $\II$ projection, but making sure that at every
stage of the analysis the configuration under study is invariant under
$\I$ and $\II$.

We shall find it more convenient to start with the BCFT describing the
D3-brane pair spanning $X^6$-$X^8$-$X^9$ directions, and
identify an $\I$ and $\II$ invariant marginal perturbation which takes
this BCFT to the BCFT describing the D1-brane pair along $X^9$ direction.
The relevant part of the BCFT at the critical radii 
before we switch on the perturbation is described by a pair of scalar fields 
$X^i \equiv X^i_L + X^i_R$ for $i = 6, 8$ and their left- and right-moving 
superpartners $\psi^i_L, \psi^i_R$. These fields satisfy Neumann boundary
conditions on the real line, given by:{\fn{ These are the boundary conditions
for NS-sector open string states. In the Ramond(R)-sector we have different 
boundary condition for $\psi^i$; it has been discussed in \cite{9808141} 
and will not be discussed here.}}
\be\label{Dirichletbc}
(X^i_L)_B = \:(X^i_R)_B \equiv \half\:X^i_B;\qquad  (\psi^i_L)_B = \:
(\psi^i_R)_B \equiv \:\psi^i_B.
\ee
Besides these there are other scalar fields and their fermionic
superpartners corresponding to the other coordinates, and also bosonic and
fermionic ghost fields. But they will play no role in our analysis.

The other part of the BCFT which will be important for our
analysis is the Chan Paton (CP) factor.
Open string
spectrum on each non-BPS D3-brane comes from two CP sectors\cite{9809111}: 
$\Id$ and $\sigma_1$ ($2\times 2$ Pauli matrix) --- these will be called
{\em{internal}} 
CP factors. For a pair of non-BPS D3-branes, there also exists a set of 
{\em{external}} CP factors : $\Id$, 
$\Sigma_1$, $\Sigma_2$, and $\Sigma_3$ ($2\times 2$ Pauli matrices). 
($\Sigma_1$, $\Sigma_2$) CP sectors correspond to open strings stretched 
between different non-BPS D3-branes,
and ($\Id$, $\Sigma_3$) correspond to open strings with both
ends on the
same
D3-brane. The {\em{internal}} and {\em{external}} CP
factors commute with each other. 

Due to the presence of $\Zop_2$ Wilson lines along $X^6$ and $X^8$ on one
of the D3-branes, the
open string states with two ends on two different D3-branes are
anti-periodic under $X^6\to X^6+2\pi R_6$ and under $X^8\to X^8+2\pi R_8$,
and hence carry half
integer units of momentum along $X^6$ and $X^8$. This can be restated as
follows: {\em the translation symmetry $X^8\to X^8+2\pi R_8$ ($X^6\to
X^6+2\pi R_6$)  under which
we
normally identify space to make $X^8$ ($X^6$) compact, acts on the open
string
states via conjugation by $\Sigma_3$}. Under this conjugation, CP sectors 
$\Id$ and
$\Sigma_3$, representing open strings with both ends on the same D3-brane,
remain invariant, but CP sectors $\Sigma_1$ and $\Sigma_2$, representing
open strings with two ends on two different branes, change sign. This is
exactly what we require.

Our next task is to determine the action of $\I$ and $\II$ on various
operators of the BCFT, so that we can identify $\I$ and $\II$ invariant
vertex operators. On the coordinates $X^6$ and $X^8$ and their fermionic
partners, the action of $\I$ and $\II$ has been specified in eqs.\refb{I}
and \refb{II}; thus we only need to determine their action on the CP
factors. We can divide this into two parts: the action on the
internal CP factor, and that on the external CP factor. The action on the
internal CP factor can be determined by taking a single D3 brane, and
identifying an appropriate 2-point coupling between a closed string state
and an open string state\cite{9812031}. Since we know the transformation
properties of
the closed string states under $\I$ and $\II$, this determines the
transformation properties of open string states under $\I$ and $\II$, and
hence their action on the internal CP factor. This procedure has been
illustrated in ref.\cite{9812031}; hence here we shall quote the final
result. Both $\I$ and $\II$ 
conjugates internal CP factor by $\sigma_3$. Thus they leave states with
internal CP factor $\Id$ unchanged, and change the sign of the states
with internal CP factor $\sigma_1$.

Let us now turn to the action of $\I$ and $\II$ on the external CP
factors. Since both $\I$ and $\II$ leave individual D3-branes unchanged,
they must leave unchanged CP factors $\Id$ and $\Sigma_3$, and hence can
at most induce a rotation about the 3-axis on the external CP factors. Since
$(\I)^2$ acts as identity on the fields, it must also act as identity on
the CP factor in order that it is an order two transformation. This leaves
us with two choices: either it acts as conjugation by $\Sigma_3$, and
changes the sign of $\Sigma_1$ and $\Sigma_2$, or it leaves all external
CP factors invariant. Both choices are allowed. After modding out the
theory by $\I$, these two choices can be shown to be
related to the choice of relative sign of $\ep^{(1)}$ and $\ep^{(2)}$ in
eqs.\refb{tw9}. From eq.\refb{econ0} we see that we need
$\ep^{(1)}=\ep^{(2)}$; this can be shown to correspond to choosing the
action of $\I$ to be trivial on all the external CP factors. For the time
being we shall proceed with this assumption, but later we shall consider
the other choice, and show
how this corresponds to choosing $\ep^{(1)}=-\ep^{(2)}$.
(The most straightforward way of seeing this
is to construct the boundary state of the D3-branes
characterized by $\ep$, $\theta_6$ and $\theta_8$, and then compute the
open string partition function by taking the inner product between two
such boundary states. But we shall follow a shorter, more intuitive path.)

Next we turn to the action of $\II$ on the external CP factors. For this
note that the action of $(\II)^2$ on various fields is given as follows:
\be \label{ei42}
(\II)^2: \qquad X^6\to X^6, \quad X^8\to X^8 + 2\pi R_8, \quad \psi^6\to
\psi^6, \quad \psi^8\to \psi^8\, .
\ee
Recall now that $X^8\to X^8+2\pi R_8$ acts as identity on all the open
string states only if it is accompanied by the conjugation of external CP
factors by $\Sigma_3$. Thus in order that $\II$ is an order two
transformation, $(\II)^2$ must conjugate external CP factors by
$\Sigma_3$. Hence $\II$ itself must conjugate the external CP factors by
a square root of $\Sigma_3$. Furthermore, we have already seen that it
should leave the external CP factors $\Id$ and $\Sigma_3$ unchanged. This
leaves us with two possible choices: conjugation by $\exp(\pm
i\pi\Sigma_3/4)$. They give equivalent results.\footnote{After modding out
the theory by $\II$, these two choices correspond to the choice of sign in
front of the part of the boundary state carrying half integer winding
along $x^8$ and $x^6$, $-$ the sectors twisted by $\II$ and $\I\II$
respectively.} For
definiteness, we shall take this to be $\exp(i\pi \Sigma_3/4)$.

With these rules we are now in a position to construct the $\I$ and $\II$
invariant vertex
operators.
At $R_6=R_8=1$, the lowest energy states of the open strings, carrying
internal CP factor $\sigma_1$, external CP factors $\Sigma_1$ or
$\Sigma_2$, and 
momenta $\pm{1\over 2}$ along $X^6$ and $X^8$, become massless. Before
requiring $\I$ and $\II$ invariance, the vertex operator of one such open
string state in the (0,0) picture\cite{FMS} is given by:
\be \label{ebef}
(\psi^6_B+\psi^8_B) \exp({i\over 2} (X^6_B+X^8_B)) \ox \sigma_1\ox
\Sigma_1\, .
\ee
The complete $\I$ and $\II$ invariant vertex operator can be constructed
by adding to it its transforms under $\I$, $\II$ and $\I\cdot\II$. This is
given by
\ben\label{tachvo3}
V^{(0)}_T & \propto & \half\,({\psi}^6_B \:+\: {\psi}^8_B)\,\Big[\,
e^{i({X}^6_B\:+\: {X}^8_B)/2}\;+ \; e^{-\,i({X}^6_B\:+\: 
{X}^8_B)/2}\,\Big]
\otimes \y\ox\z\non\\
& + & \:\half\,({\psi}^6_B \: - \: {\psi}^8_B)\,\Big[\,
e^{i({X}^6_B\: - \: {X}^8_B)/2}\;+ \; e^{-\,i({X}^6_B\: - \: 
{X}^8_B)/2}\,\Big] 
\otimes \y\ox\zz
\een 

We now note that this perturbation is identical to the one described in
ref.\cite{0003124} in the context of marginal deformation of a BPS D-brane -
$\bD$-brane system, with the only difference that there is an additional
internal CP factor $\sigma_1$ attatched to each vertex operator. Since
$\sigma_1$ commutes with all other operators, the presence of this operator 
does
not affect the analysis, and one can show following ref.\cite{0003124}
that \refb{tachvo3} corresponds to an exactly marginal
operator. One can also follow the procedure of \cite{0003124} to study how
the spectrum of open strings changes under this deformation, and show that
for a specific value of the deformation parameter the spectrum becomes
identical to that of open strings living on a pair of D1-branes (along 
$X^9$) situated at
diametrically opposite points of the torus spanned by $(X^6,X^8)$. Indeed,
part of the spectrum of open strings living on the original non-BPS D3-brane 
pair, corresponding to CP factors $\Id\ox\Id$, $\Id\ox\Sigma_3$,
$\sigma_1\ox\Sigma_1$ and $\sigma_1\ox\Sigma_2$, is identical to that
living on a D3-$\bD$3 brane pair of IIB, and these states evolve under
the marginal deformation in a manner identical to that described in
\cite{0003124}. These correspond to states living on a D1-$\bD$1 brane
pair
of IIB in the deformed theory, forming part of the expected spectrum of open
strings on a D1-brane pair of IIA. The rest of the states on the D3-brane
pair of IIA can be shown to evolve into the rest of the states on the 
D1-brane 
pair of IIA under this deformation. As the analysis is a straightforward
generalization of the one carried out in \cite{0003124}, we shall not
repeat it here.

Thus we
conclude that the marginal deformation takes the original D3-brane pair
to a D1-brane pair. The locations of the
D1-branes can be determined as follows. 
The tachyon vertex operator in the zero picture is given in
\refb{tachvo3}. Examining the $-1$ picture version of this vertex
operator,
and defining a complex tachyon field whose real and imaginary parts are
proportional to the
coefficients of $\Sigma_1$ and $\Sigma_2$ respectively in the $-1$
picture, as in
\cite{0003124}, we get 
\be \label{etachf}
T(X^6,X^8) \propto \sin({1\over 2} (X^6+X^8)) + i\sin({1\over 2}(X^6 -
X^8))\, .
\ee
This has zeroes at $X^6=X^8=0$ and at $X^6=X^8=\pi$, showing that these
are the locations of the D1-branes. This is consistent with the choice
\refb{econ0}-\refb{consistent2}, since with this choice the net twisted
sector charges are concentrated at $X^6=X^8=0$ and at $X^6=X^8=\pi$.

Now suppose we had made a different choice for the action of $\I$ on the
external CP factors, namely that it changes the sign of the external CP
factors
$\Sigma_1$ and $\Sigma_2$. In that case, \refb{tachvo3} would be replaced
by
\ben\label{tachvo33}
V^{(0)}_T & \propto & \half\,({\psi}^6_B \:+\: {\psi}^8_B)\,\Big[\,
e^{i({X}^6_B\:+\: {X}^8_B)/2}\;- \; e^{-\,i({X}^6_B\:+\: 
{X}^8_B)/2}\,\Big]
\otimes \y\ox\z\non\\
& - & \:\half\,({\psi}^6_B \: - \: {\psi}^8_B)\,\Big[\,
e^{i({X}^6_B\: - \: {X}^8_B)/2}\;- \; e^{-\,i({X}^6_B\: - \: 
{X}^8_B)/2}\,\Big] 
\otimes \y\ox\zz
\een 
This is still a marginal deformation, but would correspond to a tachyon
field configuration:
\be \label{etachf3}
T(X^6,X^8) \propto \cos({1\over 2} (X^6+X^8)) - i\cos({1\over 2}(X^6 -
X^8))\, .
\ee
Thus now it has zeroes at $(X^6=0,X^8=\pi)$ and at $(X^6=\pi, X^8=0)$.
This is where the final state D1-branes will be, and hence this is where
the twisted sector charges will be concentrated. By examining
eq.\refb{tw9} we see that this corresponds to the choice
\be \label{enewch}
\ep^{(1)}=-\ep^{(2)}\, ,
\ee
together with eqs.\refb{consistent1} and \refb{consistent2}. This shows
that the choice of how $\I$ acts on external CP factors is correlated with
the choice of relative signs of $\ep^{(1)}$ and $\ep^{(2)}$, and that for
our system, $\I$ acts trivially on the external CP factors.

As in ref.\cite{0003124} one can deform the system away from the critical
radii, and show that the final system continues to describe a D1-brane
pair. This establishes that the D1-brane pair indeed decays into a
D3-brane pair (and vice versa) across the region of stability $R_6R_8=1$.

\bigskip

\noi{\bf Acknowledgement}:
We would like to thank Debashis Ghoshal, Dileep Jatkar, Suresh
Govindarajan,
Sunil Mukhi and B. Stefansky for useful discussions. J.M. would also like
to thank
The Department of Theoretical 
Physics, TIFR, India, for their hospitality, where part of this work was 
done.  A.S. would like to thank the New High Energy Theory Center at
Rutgers University, Center for Theoretical Physics at MIT, and the Erwin
Schroedinger Institute, Vienna for hospitality where part of this work was
done.

\end{document}